\documentclass[12pt]{article}

\usepackage{latexsym,epsfig,amssymb,euscript,amsmath,verbatim}

 \def\be{\begin{equation}} \def\ee{\end{equation}}
\def\bea{\begin{eqnarray}} \def\eea{\end{eqnarray}}

\begin{document}

\pagestyle{empty}
\rightline{SISSA-38/2010/EP}

\rightline{ULB-TH/10-21}
\vspace{0.8cm}
\begin{center}
{\LARGE{\bf Unscreening the Gaugino Mass \\
\vspace{7mm}
with Chiral Messengers}}

\vskip 20pt
 {\large{Riccardo Argurio$^{1}$,
Matteo Bertolini$^{2,3}$, Gabriele Ferretti$^{4}$ \vskip 3pt and  Alberto Mariotti$^{5}$ \\[5mm]}}
{\small{{}$^1$ Physique Th\'eorique et Math\'ematique and International Solvay
Institutes \\
\vspace*{-2pt}  Universit\'e Libre de Bruxelles, C.P. 231, 1050
Bruxelles, Belgium\\
\medskip
{}$^2$ SISSA and INFN - Sezione di Trieste\\
\vspace*{-2pt} Via Bonomea 265; I 34136 Trieste, Italy\\
\medskip
{}$^3$ International Centre for Theoretical Physics (ICTP)\\
\vspace*{-2pt} Strada Costiera 11; I 34014 Trieste, Italy\\
\medskip
{}$^4$  Department of Fundamental Physics \\
\vspace*{-2pt} Chalmers University of Technology, 412 96 G\"oteborg, Sweden
\\
\medskip
{}$^5$ Theoretische Natuurkunde and International Solvay Institutes \\
\vspace*{-2pt}
Vrije Universiteit Brussel, Pleinlaan 2, B-1050 Brussels, Belgium}}\\
\medskip

\medskip

\medskip

\medskip

{\bf Abstract}
\vskip 20pt
\begin{minipage}[h]{16.0cm}
Gaugino screening, the absence of next-to-leading order corrections to gaugino masses,
is a generic feature of gauge mediation models of
supersymmetry breaking. We show that in a specific class of models,
known as semi-direct gauge mediation, it is
possible to avoid gaugino screening by allowing for a chiral
messenger sector. Messengers then acquire a mass at some scale, for instance by higgsing or
by some auxiliary strong coupling dynamics. We implement
this idea in a simple model which we work out explicitly.
\end{minipage}
\end{center}

\newpage




\setcounter{page}{1} \pagestyle{plain}

\renewcommand{\thefootnote}{\arabic{footnote}} \setcounter{footnote}{0}



\section{Introduction}

In the framework of gauge mediation of supersymmetry (SUSY) breaking
\cite{Giudice:1998bp}, gaugino mass screening refers to the fact that
next-to-leading order radiative corrections to the gaugino mass are absent
\cite{ArkaniHamed:1998kj}. When the visible
gaugino mass vanishes at leading order, as in semi-direct gauge
mediation (SDGM) \cite{Randall:1996zi,Seiberg:2008qj},
the screening implies that the gaugino mass will only arise at next-to-next-to-leading
order and hence be severely suppressed with respect to sfermions masses. While such
strong hierarchy would fit into
a split supersymmetry scenario \cite{ArkaniHamed:2004fb}, one might wonder
whether there are ways to avoid gaugino screening in gauge
mediation, in the first place.

\begin{figure}[ht]
\centering
\includegraphics[width=0.75\textwidth]{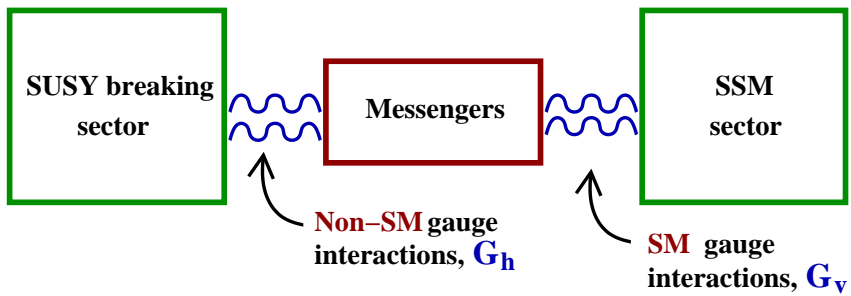}
\caption{\small A cartoon of semi-direct gauge
mediation. The gauge group $G_h$ is singled-out within the hidden
sector as the subgroup to which the messenger fields couple. $G_v$ is the gauge group of the visible sector.
\label{semi}}
\end{figure}

The object of this note is to present a SDGM set up where the
phenomenon of gaugino mass screening does not take place. SDGM is
a class of gauge mediation models where the messengers interact
with the hidden sector only through (non-SM) gauge interactions
but, unlike direct gauge mediation, they do not participate to
the hidden sector supersymmetry breaking dynamics. A pictorial
representation of SDGM is reported in figure \ref{semi}.

\vskip 7pt

The idea to avoid gaugino screening in SDGM is very simple: it suffices to
allow for a chiral messenger sector, in the sense that gauge symmetries prevent the presence of an explicit
mass term in the superpotential for the messenger fields. Eventually, the
messengers will acquire a mass (e.g.~by higgsing) and disappear from the low energy spectrum. However, if there is a sufficient
range for RG evolution above this scale the visible gaugino will indeed acquire a non-vanishing mass at
next-to-leading order. In spirit this is very close to (and
indeed inspired by) what is called $Z'$ mediation
\cite{Langacker:2007ac}, where the role of
the messengers is played directly by the MSSM matter.

In what follows we discuss the evasion of the gaugino mass screening argument
both from the point of view of its original discussion in terms of
wave function renormalization \cite{ArkaniHamed:1998kj}, in section \ref{wavefunction}, and by direct evaluation of
Feynman diagrams \cite{Argurio:2009ge}, in section \ref{evaluation}.
In section \ref{model} we implement our basic idea within a concrete model based on a quiver gauge
theory, which can arise from D-branes at singularities, and provide additional details on the mechanism
of gaugino unscreening. We end in section \ref{pheno} by discussing possible phenomenological implementations of
our model, also giving estimates for the gaugino and sfermion masses.

\section{Gaugino (un)screening from wave function \\ renormalization}
\label{wavefunction}

As we mentioned in the introduction, what goes under the name of gaugino mass screening is the observation
that when gauge mediation of SUSY breaking is operated by messengers,
next-to-leading order corrections to the gaugino mass cancel each
other. In the context in which one obtains soft masses by promoting MSSM
wave function renormalization factors to (spurionic) superfields,
the argument goes as follows \cite{ArkaniHamed:1998kj}.

The expression for the running physical (real) gauge coupling $R(\mu)$ is given
by (for ease of comparison, we stick to the notation of
\cite{ArkaniHamed:1998kj})
\be
R(\mu) = S(\mu) + S(\mu)^\dagger + \frac{T_G}{8\pi^2} \log (S(\mu) +
S(\mu)^\dagger) -\sum_r \frac{T_r}{8\pi^2} \log Z_r(\mu)~,
\ee
where $S(\mu)$ is the
holomorphic coupling and $Z_r(\mu)$ the wave function
renormalization of matter fields. The sum over $r$ is on all representations (of index $T_r$) of
matter fields charged under the gauge group which are present below
the scale $\mu$. We recall that the presence of the logarithmic terms
is in order to compensate the unphysical rescaling symmetry of the
holomorphic coupling $S(\mu)$.

The running of $R(\mu)$ will experience a threshold at the scale at
which the messengers stop contributing (when going towards the IR). We
will call this scale $\mu_X$. The RG running will be different above
and below this scale, essentially because of the presence, above
$\mu_X$, of an extra term depending on the wave function
renormalization of the messengers. Eventually, what we find for the
gauge coupling for $\mu<\mu_X<\mu_0$ is
\bea
\label{runnc}
R(\mu)& =& R(\mu_0) + \frac{b_0}{16\pi^2} \log \frac{\mu^2}{\mu_0^2}
+ \frac{T_G}{8\pi^2} \log \frac{\mathrm{Re} S(\mu)}{\mathrm{Re} S(\mu_0)}
 -\sum_r \frac{T_r}{8\pi^2} \log \frac{Z_r(\mu)}{Z_r(\mu_0)} \nonumber
 \\
& & \qquad -  \frac{T_M}{16\pi^2} \log \frac{\mu_X^2}{\mu_0^2}
- \frac{T_M}{8\pi^2} \log \frac{Z_M(\mu_X)}{Z_M(\mu_0)}~,
\eea
where $T_M$ is essentially the number of messengers, and $Z_M$ their
wave function renormalization. The constant $b_0 = 3 T_G - \sum_r T_r$ is the coefficient
of the one-loop beta function below the scale $\mu_X$.
The leading order contribution to the gaugino mass comes from
replacing $\mu_X$ by its tree level value $X$ and then promoting it to
a spurion $X+\theta^2 F$. Next-to-leading order corrections should come
from corrections to the values of $\mu_X$ and $Z_M$. The fact that next-to-leading order corrections
vanish derives precisely from the fact that the correct expression for
$\mu_X$ takes into account wave function renormalization
\be
\label{scalem}
\mu_X = \frac{X}{Z_M(\mu_X)}~.
\ee
We are then left with
\bea
R(\mu) &=& R(\mu_0) + \frac{b_0}{16\pi^2} \log \frac{\mu^2}{\mu_0^2}
+ \frac{T_G}{8\pi^2} \log \frac{\mathrm{Re} S(\mu)}{\mathrm{Re} S(\mu_0)}
 -\sum_r \frac{T_r}{8\pi^2} \log \frac{Z_r(\mu)}{Z_r(\mu_0)} \nonumber
 \\ & & \qquad \qquad \qquad
-  \frac{T_M}{16\pi^2} \log \frac{X^2}{\mu_0^2Z_M(\mu_0)^2}~.
\eea
While the last term in the equation above provides the leading order contribution to gaugino mass, there are no
next-to-leading order corrections, as anticipated.

In a SDGM set up \cite{Randall:1996zi,Seiberg:2008qj} gaugino screening has dramatic
consequences since the messenger mass $X$ does {\em not} acquire an F-term at tree level,
so there are not even leading corrections to the (visible)
gaugino mass, which is then zero up until next-to-next-to-leading order
\cite{ArkaniHamed:1998kj}. This result concerns the contribution to
the gaugino mass at linear order in $F$ and at all orders in any hidden gauge
or self-interaction coupling, but it does not exclude contributions of
higher order in $F$. However, those are also all vanishing at leading
order in the hidden gauge coupling \cite{Argurio:2009ge}.

Now, the question is whether it is possible to evade this argument, which seems quite
general and robust. The answer is surprisingly simple: let us allow for a chiral messenger spectrum and
consider the physical gauge coupling at some scale $\mu < \mu_0$ (the beta function coefficient is now
$b'_0 = b_0 - T_M$)
\bea
\label{runc}
R(\mu)& =& R(\mu_0) + \frac{b'_0}{16\pi^2} \log \frac{\mu^2}{\mu_0^2}
+ \frac{T_G}{8\pi^2} \log \frac{\mathrm{Re} S(\mu)}{\mathrm{Re} S(\mu_0)}
 -\sum_r \frac{T_r}{8\pi^2} \log \frac{Z_r(\mu)}{Z_r(\mu_0)} \nonumber
 \\
& & \qquad \qquad \qquad
- \frac{T_M}{8\pi^2} \log \frac{Z_M(\mu)}{Z_M(\mu_0)}~.
\eea
Let us now suppose that the messenger wave function $Z_M(\mu)$
experiences a SUSY breaking threshold at some scale between $\mu$ and $\mu_0$.
For instance, the messengers could couple to a hidden gauge group,
whose gaugino obtains a mass. The latter can be seen as a spurionic
F-term to the holomorphic hidden gauge coupling. Having a chiral messenger spectrum,
hidden gauge radiative corrections to the wave function renormalization of the
messengers, being now unbalanced, will propagate an F-term down to the visible
gauge coupling function, and gaugino screening would then not occur. This is
actually exactly what happens in $Z'$ mediation, except that the messenger's role
is played by the MSSM chiral matter.

Of course, having a chiral messenger spectrum cannot be ultimate the solution, since massless
messengers are not acceptable, phenomenologically. Hence, they must
acquire a mass at some stage, for instance by
higgsing. Well below this mass, the RG evolution is as in the
case of massive messengers. However, if the messengers mass is
sufficiently smaller than the scale of SUSY breaking, there is enough RG
evolution between the two to produce a non-vanishing visible gaugino
mass.

In essence, before higgsing the messengers will be in a
chiral representation of the gauge groups. By
consequence, they will have chiral couplings to the hidden and visible
gauge groups, the wave function renormalization will be different for
the two chiralities of the messengers, and the scale matching (\ref{scalem}) crucial
in obtaining the cancellation at next-to-leading order cannot be
done. Below the higgsing scale, instead, one recovers a non-chiral
spectrum and therefore the RG flow does not feed the visible gaugino mass
anymore.


\section{Direct evaluation of the gaugino mass}
\label{evaluation}

Here we give a different, more direct argument in favour of gaugino
mass unscreening with chiral messengers.

\begin{figure}[ht]
\centering
\includegraphics[width=0.85\textwidth]{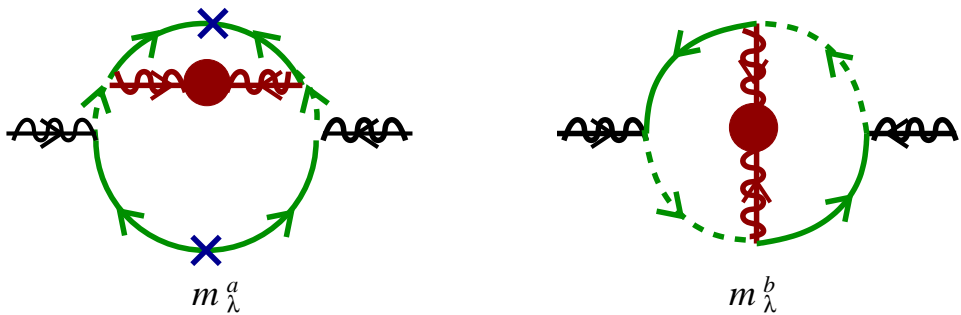}
\caption{\small The diagrams contributing to the gaugino mass. The external line corresponds to
the visible gaugino $\lambda$, the internal line with the blob attached corresponds to the propagator
of the hidden gaugino (the blob encodes the exact hidden sector non-supersymmetric correction to the
propagator), while all other internal lines correspond to messenger fields. The left diagram has two
(supersymmetric) mass insertions, each one represented by a cross on the corresponding messenger fermionic line.
\label{gaugino}}
\end{figure}

We recall from \cite{Argurio:2009ge} that
in SDGM there are only two types of diagrams contributing to the gaugino mass, as displayed
in Figure \ref{gaugino}.

\vskip 5pt

\vskip -6pt
The result of gaugino mass screening (at leading order in the hidden
gauge coupling, but to all orders in $F$) comes about by noticing that
the two diagrams cancel each other exactly at zero momentum, and independently of the
SUSY breaking current insertion on the hidden gaugino (chiral)
propagator \cite{Argurio:2009ge}.

When messengers are chiral, and hence massless, the cancellation no
longer holds for a trivial reason: one cannot write the first
diagram, $m^a_\lambda$,
since it would involve mass insertions on the fermionic messenger
lines. Then, the visible gaugino mass is non zero and given by the
massless limit of the second diagram, with messengers of only a
single chirality running in the loop. A very similar diagram appears indeed
in the context of $Z'$ mediation \cite{Langacker:2007ac}.

It is quite straightforward to evaluate explicitly this diagram. From each chiral
messenger, in the limit in which we can consider it massless, we obtain a contribution given by
(suppressing group theory factors)
\be
\label{inovis}
m_{\lambda} = 4 g_v^2 g_h^2 \int \frac{d^4k}{(2\pi)^4} \int
\frac{d^4l}{(2\pi)^4} \frac{l\cdot (l-k) }{ l^4 (l-k)^4}
\frac{g_h^2 M B(k^2/M^2) }{k^2+(g_h^2 M B(k^2/M^2))^2}~.
\ee
In the above, $g_v$ and $g_h$ are the couplings of the
visible and the hidden gauge groups, respectively, in which the messengers are
bifundamentals, $M$ is a scale related to the SUSY breaking dynamics
in the hidden sector, while $B(k^2/M^2)$ is the chiral
correlator of the hidden sector fermionic current that determines the
mass of the hidden gaugino $\lambda_h$, see \cite{Argurio:2009ge,Meade:2008wd}.
Note that we have resummed the
hidden gaugino chiral propagator in order to avoid IR divergences.

First of all we evaluate by standard techniques the
kernel
\be
\int \frac{d^4l}{(2\pi)^4} \frac{l\cdot (l-k) }{ l^4 (l-k)^4} =
\frac{1}{(4\pi)^2} \frac{1}{k^2}~.
\ee
Note that this is the correct $m\to 0$ limit of the kernel that one
writes for messengers of mass $m$ (after factoring out a power of $m^2$),
and that was computed in \cite{Argurio:2009ge}.

We can now use this result to compute the visible gaugino mass.
For definiteness, we approximate $B(k^2/M^2)$ by a step function, and
we take the hidden gaugino mass to be $m_{\lambda_h}=g_h^2MB(0)$ as we set ourselves in the regime $m_{\lambda_h}\ll M$. We get
from eq.~(\ref{inovis})
\bea
m_{\lambda} & = & 4 \frac{g_v^2 g_h^2}{(4\pi)^2}
\int \frac{d^4k}{(2\pi)^4} \frac{1}{k^2} \frac{g_h^2 M B(k^2/M^2)
}{k^2+(g_h^2 M B(k^2/M^2))^2}  \nonumber \\
 & = & 4 \frac{\alpha_v}{4\pi} \frac{\alpha_h}{4\pi} \int_0^{M^2} dk^2
\frac{m_{\lambda_h}}{k^2 + m_{\lambda_h}^2} 
 \sim \frac{\alpha_v}{4\pi} \frac{\alpha_h}{4\pi} m_{\lambda_h}
\log \frac{M^2}{m_{\lambda_h}^2}~. \label{mlh} \eea
As already noticed, the messengers will eventually get a mass by some
model-dependent dynamical mechanism (e.g. higgsing, confinement). However, assuming
that the dynamical mass scale scale is much smaller than the hidden gaugino
mass, the above expression will only have
negligible corrections. Even in the case where the
messengers' acquired mass is of the same order of $m_{\lambda_h}$,
but still much smaller than $M$, it can be shown that the
expression above will be corrected at most by an ${\cal O}(1)$
factor.

\section{A model of chiral messengers}
\label{model}

In this section we present a model that implements the ideas developed above.
Our goal is not to present a complete phenomenologically viable model,
but to show that the idea discussed in the previous sections can find a concrete realization.

\begin{figure}[ht]
\centering
\includegraphics[width=0.40\textwidth]{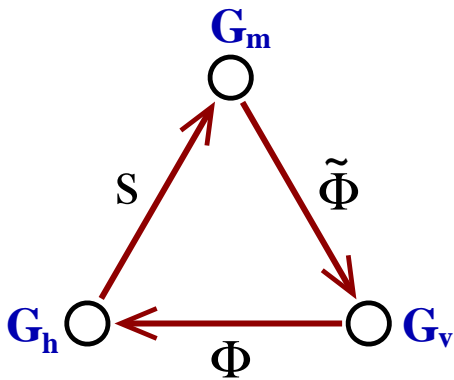}
\caption{\small The quiver gauge theory arising at a $dP_3$ Calabi-Yau singularity
describing the messenger sector and its interactions. Visible matter
fields are attached to the group $G_v$, and can
be engineered in terms of flavor D7-branes at the singularity. The
SUSY breaking dynamics couples instead only to $G_h$.
\label{dp3}}
\end{figure}

The several gauge groups and chiral superfields needed in a model of
SDGM can
be easily encoded in a quiver gauge theory that can actually be found among those arising
from D-branes at Calabi-Yau singularities \cite{Argurio:2009pz}. The specific model we
consider here can be obtained, for instance, by considering fractional D3-branes at a del Pezzo 3
singularity, and is depicted in figure \ref{dp3}.

There are three gauge groups, whose ranks should be one and the same to avoid gauge
anomalies. In the following we assume $G_v=G_h=G_m=SU(5)$, having in mind applications to GUT
theories. There are three different bifundamental superfields: the chiral superfield $\Phi$ charged under $G_v$ and
$G_h$, the chiral superfield $\tilde \Phi$ charged under $G_m$ and $G_v$, and an extra
superfield $S$ charged under $G_h$ and $G_m$. The superfields $\Phi$ and $\tilde \Phi$ transform in the
$5$ resp. the $\bar 5$ of the GUT group $G_v$.
In addition, there is a (unique) superpotential term
\be
W= y S \tilde \Phi \Phi~.
\ee
Of course, besides these fields there will be other
chiral superfields charged only under the visible gauge group $G_v$, forming the chiral matter of the visible sector.
Similarly, there will be extra dynamical fields affecting only the hidden sector $G_h$, that will give rise
to supersymmetry breaking but whose detailed dynamics will not be addressed here.
The presence of these fields will always be understood in the following but we will concentrate only on those
fields charged under at least two of the groups. The third gauge group, $G_m$, is needed to make the whole
theory free of gauge anomalies.

Notice in passing that one could also reduce the number of messengers as seen by the visible sector by replacing
$G_h$ and $G_m$ with some lower rank group and attaching to
them enough extra matter to cancel their cubic anomalies
(or, in the case of $SU(2)$ the global anomaly arising from an odd number of fields).

The transition from the chiral messenger model at higher energies to a
model where the messengers are eventually massive is done by giving a
diagonal VEV $v$ to $S$. Once $S$ has a VEV, the messengers obtain a supersymmetric mass equal to $y v$, and $G_h$ and
$G_m$ are higgsed to a diagonal $SU(5)$.

In the absence of SUSY
breaking the off-diagonal combination of the
two gauginos $\lambda_h$ and
$\lambda_m$ would get a Dirac mass by mixing with the fermion in $S$,
while the diagonal gaugino would stay massless.

The story changes if SUSY breaking is present and affects $G_h$. Then,
before higgsing, the $G_h$ gaugino $\lambda_h$ already has a
(Majorana) mass. After higgsing it will mix (negligibly
if we assume $g_hv, g_mv\ll m_{\lambda_h}$) with the $G_m$
gaugino $\lambda_m$, however the messengers $\Phi$ and $\tilde \Phi$,
even if massive, will still couple to a different gaugino. Hence, to leading order, the contribution to the visible gaugino mass
will be the one discussed in the previous section.

Let us see this is in a bit more detail. Our superfields are
$S=v+\theta \sigma +\dots$, $\Phi=\phi+\theta \psi +\dots$ and
$\tilde \Phi=\tilde \phi+\theta \tilde \psi +\dots$.
After higgsing, the terms in the SUSY Lagrangian bilinear in the
relevant fermions are
\be
{\cal L} \supset
{\cal L}_{ferm}=
i \sqrt{2}g \phi^* \psi \lambda_h - i\sqrt{2} g v \sigma \lambda_h + i \sqrt{2}gv
\sigma \lambda_m - i\sqrt{2} g \tilde \phi^* \tilde \psi \lambda_m + y v \tilde
\psi \psi~.
\ee
For convenience, we have set the two couplings $g_h$ and $g_m$ to the
same value $g$ and have once again dropped the group theory factors.
To the above Lagrangian, we have to add the SUSY
breaking mass for the hidden gaugino
\be
\label{explicit} {\cal L}' = \mbox{$\frac{1}{2}$} m_{\lambda_h}
\lambda_h \lambda_h~. \ee The lagrangian ${\cal{L}}_{ferm}+{\cal
L}'$ characterizes the fermionic sector of the theory. Note that
the higgsing scale $g v$ can be different from the messenger mass
scale $y v$. Moreover, we have not yet assumed any specific relation between
the different scales $g v$, $yv$ and $m_{\lambda_h}$.

We are interested in computing the contributions to the visible gaugino mass.
The diagrams are the ones in figure \ref{gaugino}. Observe that the visible gaugino
couples to the messengers, whose fermions have a Dirac mass in ${\cal L}_{ferm}$.
The messengers then couple to the hidden and messenger gauginos
$\lambda_h$ and $\lambda_m$.
The shorter way to perform the computation is to invert the quadratic part of the Lagrangian
for the fermions ($\lambda_h$, $\lambda_m$, $\sigma$) and extract the two point functions
\bea
{\cal B}_{hh} &\equiv&
\langle \lambda_h \lambda_h \rangle=
\frac{m_{\lambda_h} (k^2+ 2g^2 v^2 )^2}{
k^2(k^2+4g^2 v^2)^2 + m_{\lambda_h}^2(k^2+ 2g^2 v^2 )^2}
\nonumber \\
\label{twopoints}
{\cal B}_{mm} &\equiv&
\langle \lambda_m \lambda_m \rangle=
\frac{4g^4 v^4 m_{\lambda_h}}{
k^2(k^2+4g^2 v^2)^2 + m_{\lambda_h}^2(k^2+ 2g^2 v^2 )^2}
\\
{\cal B}_{hm} &\equiv&
\langle \lambda_h \lambda_m \rangle=
\frac{2g^2 v^2 m_{\lambda_h} (k^2 + 2g^2 v^2 )}{
k^2(k^2+4g^2 v^2)^2 + m_{\lambda_h}^2(k^2+ 2g^2 v^2 )^2}~.
\nonumber
\eea
Note that all of them vanish in the supersymmetric limit $m_{\lambda_h}=0$.

The two point functions computed above enter into the computation of the diagrams of figure \ref{gaugino}
as the blobs in the internal lines. It is easy to see that in the diagram to the left of figure 2, the
internal gaugino line is $\langle \lambda_h \lambda_m\rangle$, while
there are two diagrams corresponding to the one on the right, one with
a $\langle \lambda_h \lambda_h\rangle$ line and the other with a
$\langle \lambda_m \lambda_m\rangle$ line.

With a computation similar to the one in \cite{Argurio:2009ge}
we can obtain the resulting contribution to the visible gaugino mass
\be
\label{gaugino1}
m_{\lambda}=8 g_v^2 g^2
\int \frac{d^4 k}{(2\pi)^4}
\left(
L_a(k^2, (yv)^2) {\cal B}_{hm}+L_b(k^2,(yv)^2)\frac{ {\cal B}_{hh}+{\cal B}_{mm}}{2}
\right)~,
\ee
where $L_a$ and $L_b$ have been computed in \cite{Argurio:2009ge}, and we recall
that $L_a=-L_b$. The explicit expression (rescaled by $1/m^2$ with respect to \cite{Argurio:2009ge}, where in the
present case $m=yv$) is
\be
L_b(k^2,m^2) = \frac{1}{2(4\pi)^2}\left( \frac{1}{k^2}+\frac{1}{k^2+4m^2}
- \frac{16m^4}{[k^2(k^2+4m^2)]^{3/2}} \mathrm{arctanh}
\sqrt{\frac{k^2}{k^2+4m^2}}\right)~. \label{kernelm}
\ee
The final expression for the gaugino mass hence reads
\be
\label{integralone}
m_{\lambda}=4 g_v^2 g^2
\int \frac{d^4 k}{(2\pi)^4}
 L_b(k^2,(yv)^2) \left( {\cal B}_{hh}+{\cal B}_{mm}- 2 {\cal B}_{hm} \right)~.
\ee
Recalling the explicit form of the two point functions (\ref{twopoints}), one sees that the gaugino mass
is logarithmically UV divergent. This is expected since we introduced an explicit soft supersymmetry breaking
term (\ref{explicit}).  The natural ultraviolet cut off is the supersymmetry breaking scale $M$.

Unfortunately we cannot compute the integral (\ref{integralone}) exactly.
However, we can study some interesting limits.
The combination multiplying the kernel $L_b$ in eq.~(\ref{integralone}) is
\be
\label{combo}
{\cal B}_{hh}+{\cal B}_{mm}-
2 {\cal B}_{hm}=
\frac{m_{\lambda_h} k^4}{
k^2(k^2+4g^2 v^2)^2 + m_{\lambda_h}^2(k^2+ 2g^2 v^2 )^2}~.
\ee
First, the supersymmetric case is recovered for
$m_{\lambda_h} \to 0$. In this limit the combination (\ref{combo}) and in particular each of the
two point functions in (\ref{twopoints}) vanishes.

Gaugino screening can be recovered letting $ g v \gg  M$ with arbitrary $y v$.
This corresponds to a higgsing at very high scale. The resulting effective
theory is like the one studied in \cite{Argurio:2009ge}.
In this limit the integral (\ref{integralone}) is vanishing as $\sim M^4/g^4v^4$.

In any other limit, and in particular as long as $M$ is the highest scale in the model, it is obvious
that (\ref{combo}) does not vanish and that there will be a contribution to the visible gaugino mass at this order.

We can consider the regime where $gv \ll m_{\lambda_h}$ and
also $yv \ll m_{\lambda_h}$.
In this limit the higgsing can be considered as a subdominant effect with respect to SUSY
breaking, at least as far as the fermionic sector is concerned. This can be seen
by analyzing the leading contribution for the two point functions
\be
\label{limitino}
{\cal B}_{hh}=\frac{m_{\lambda_h}}{ k^2+m_{\lambda_h}^2}
\quad , \quad
{\cal B}_{mm}=
\frac{m_{\lambda_h} 4g^4 v^4}{k^4 (k^2+m_{\lambda_h}^2)}
\quad , \quad
{\cal B}_{hm}=
\frac{m_{\lambda_h}2 g^2 v^2}{k^2 (k^2+m_{\lambda_h}^2)}~.
\ee
At leading order only ${\cal B}_{hh}$ contributes to the integral (\ref{integralone}),
and we can compute it as
\be
\label{gauginofinale}
m_\lambda=
\frac{4 g_v^2 g^2 m_{\lambda_h}}{(4 \pi)^4}
\int_0^{M^2} d k^2
\frac{1}{k^2+ m_{\lambda_h}^2 } \sim
\frac{4 g_v^2 g^2 m_{\lambda_h}}{(4 \pi)^4}
\log \frac{M^2}{ m_{\lambda_h}^2}~,
\ee
which is the same result as in section \ref{evaluation}.
The first corrections to this expression can be easily computed expanding the kernel
(\ref{kernelm}) and the combination (\ref{combo}) and performing the integral.
They scale like $(yv)^2/m_{\lambda_h}$ and
$(gv)^2/m_{\lambda_h}$.
The analytic result (\ref{gauginofinale}) is then robust for
$ gv  \ll m_{\lambda_h}$ and $yv \ll m_{\lambda_h}$.

In the most general case, $gv, yv, m_{\lambda_h} \ll M$,
we can perform the integral (\ref{integralone}) numerically.
One eventually obtains a result which is
essentially of the form (\ref{gauginofinale}) with the
$\log$ factor
replaced by a smaller ${\cal O}(1)$ factor.

This calculable model makes it clear that unscreening of the visible gaugino mass
is possible, and actually still holds even when taking into account
that the messengers eventually do get a mass. What is important of
course is that there is ultimately a sizable hierarchy between the scale of higgsing, $v$ and the SUSY breaking scale $M$.

One might be turned off by the fact that the scale of the VEV $v$,
which has to be small with respect to the SUSY breaking scale $M$, is
essentially introduced by hand.\footnote{
The VEV $v$ is essentially a (Goldstone) modulus. One can try to fix it
in several ways, the most straightforward being promoting the $SU(5)$
groups to $U(5)$ (compensating the mixed anomalies by a
Green-Schwarz-like mechanism) and then turning on a FI parameter for
the off-diagonal $U(1)_{h-m}$.}
In fact, our model presents itself an alternative
possibility. In the absence of a higgsing, the
gauge group $G_m$ will confine at a scale $\Lambda_m$ which we can
naturally take to be smaller than or of the order of $m_{\lambda_h}$. At energies well above $\Lambda_m$
the theory is chiral and the arguments of the previous section, with massless chiral messengers,
apply.
At energies below $\Lambda_m$ the theory confines and we turn to an
effective description.
The group $G_m$ has 5 colors and 5
flavors, hence we can set ourselves on the baryonic branch of the
moduli space, where the meson superfields have zero VEVs. The mesons
will act as composite messengers
\be
\tilde \Phi_\mathrm{comp} = \frac{1}{\Lambda_m}S\tilde \Phi~.
\ee
and both messengers, $\Phi$ and $\tilde \Phi_\mathrm{comp}$, will get a mass
of order $y \Lambda_m$. We are now in a situation of semi-direct gauge mediation
in all similar to the one described in
\cite{Argurio:2009ge}, so we expect to have no contribution to the
visible gaugino mass below this scale.

Clearly in this strongly coupled model the transition around
$\Lambda_m$ is less under control, but the gross features of the visible
soft spectrum should be quite similar to the previous case.

\section{Sfermion masses and phenomenology}
\label{pheno}

We have not yet discussed sfermion masses. This is a
model-dependent issue and chiral messenger models as the ones we have
discussed here have a potential problem, in this respect.

In SDGM one gets, generically, a non-supersymmetric contribution to
the messenger mass squared which provides a non-vanishing supertrace. If
this contribution is negative it can overwhelm the supersymmetric messenger mass and make the messengers
tachyonic. If on the other hand the contribution is positive, there can instead be problems
with the sfermions of the visible sector that will generically
acquire negative squared masses, since the latter is proportional to minus
the supertrace of the messenger mass matrix squared \cite{Argurio:2009ge}
(see also \cite{Poppitz:1996xw}).

Notice that the contribution to the supertrace depends on the hidden sector current correlators
$C_s^h$ \cite{Argurio:2009ge,Meade:2008wd}, which are thus not directly related to the correlator
$B^h$ entering the expressions for the gaugino mass.

Let us discuss the two above possibilities in turn. If the messenger supertrace is positive, the sfermions
are all tachyonic. In this scenario, all we can do is to find a mechanism to suppress the sfermion masses. There
are several
such mechanisms, for instance sequestering \cite{Randall:1998uk} (see also \cite{Inoue:1991rk}) by a large extra
dimension \cite{Mirabelli:1997aj}--\cite{McGarrie:2010kh} or by coupling to a conformal sector
\cite{Luty:2001zv,Schmaltz:2006qs}, deconstruction \cite{Csaki:2001em}, or holographic gauge
mediation \cite{Benini:2009ff,McGuirk:2009am}. All such models eventually lead to gaugino mediation,
where the sfermion squared masses are positive and are generated by RG flow below the scale of the visible gaugino.
Hence, in the framework of gaugino mediation, we could use our model of chiral SDGM to generate the gaugino mass
in the first place. Otherwise, the phenomenology is the same as the one of a generic gaugino mediated scenario.

In the other scenario, where the messenger supertrace is negative, the sfermion squared masses are positive
but the messengers must have a sufficiently positive SUSY mass to compensate for the negative
supertrace. In this case, some more tuning is needed. Indeed, we need to ensure that the
messengers are not tachyonic by enforcing the bound
\be
yv> \frac{\alpha_h}{4\pi} M~.
\ee
We can now compare gaugino versus generic sfermion masses $m_{sf}$ (again not paying attention to the group theory
factors that can be easily reinstated). The computation of the
sfermion masses is unaffected by the fact that the messengers are chiral and we can borrow the result
from \cite{Argurio:2009ge} where we set to $yv$ the messengers mass
\be
m^2_{sf} \sim \left(\frac{\alpha_v}{4\pi}\right)^2
\left(\frac{\alpha_h}{4\pi}\right)^2  M^2 \log \frac{M^2}{(yv)^2}~,
\ee
where the limit $yv\ll M$ is understood. This is to be compared with the visible gaugino
mass where we have
\be
\label{gauginov}
m_{\lambda} \sim \left(\frac{\alpha_v}{4\pi}\right)
\left(\frac{\alpha_h}{4\pi}\right) m_{\lambda_h}  \log
\frac{M^2}{m_{\lambda_h}^2}~ .
\ee
Assuming that the $\log$ factors and other corrections are of order unity, we see that the ratio between
gaugino and sfermion masses is given by%
\footnote{Here and above we have assumed that the hidden sector correlators are still perturbative,
hence the factors of $1/4\pi$. If they arise directly from strongly coupled dynamics one should
omit this extra factor.}
\be
\label{gauginov2}
\frac{m_{\lambda} }{m_{sf}} \sim \frac{m_{\lambda_h}}{M} \sim \frac{\alpha_h}{4\pi}~.
\ee
It is then possible to achieve a not too split visible spectrum if
$\frac{\alpha_h}{4\pi}$ is not too small. A reasonable ordering of scales
that one could aim for is the following
\be
 m_{\lambda} < m_{sf} < m_{\lambda_h},gv < yv < M~,
\ee
with one or two orders of magnitude between each scale. For instance, we could
take $\alpha_h/4\pi \sim 10^{-2}$. If we start then with $m_\lambda \sim 10^2$ GeV,
we get $m_{sf}\sim 10^4 $ GeV, $m_{\lambda_h}\sim g_h v \sim 10^5 $ GeV, $yv \sim 10^6$
GeV and finally $M\sim 10^7$ GeV. (The values given for the visible sector particles
are to be considered as boundary conditions for the MSSM RG flow, as usual.) Note that
we have to require that the coupling $y$ is at the edge of perturbativity,
$y^2/(4\pi)^2 \lesssim 1$. We conclude that even in the case of negative supertrace
it is possible to produce a soft spectrum which is not too hierarchical, although in
a small region of the parameter space.


\subsection*{Acknowledgments}
We would like to thank L.~Matos for discussions and email
exchanges, and A.~Romanino for helpful comments on a
preliminary version of the draft. In addition, R.A.~would like to
thank D.~Green and Y.~Ookouchi for informative discussions on
related topics. The research of R.A. is supported in part by
IISN-Belgium (conventions 4.4511.06, 4.4505.86 and 4.4514.08).
R.A. is a Research Associate of the Fonds de la Recherche
Scientifique--F.N.R.S. (Belgium). The research of G.F. is
supported in part by the Swedish Research Council
(Vetenskapsr{\aa}det) contract 80409701. A.M. is a Postdoctoral
Researcher of FWO-Vlaanderen. A.M. is also supported in part by
FWO-Vlaanderen through project G.0428.06. R.A. and A.M. are
supported in part by the Belgian Federal Science Policy Office
through the Interuniversity Attraction Pole IAP VI/11.



\end{document}